\documentclass[prd,amsmath,amssymb,showpacs,superscriptaddress,preprint,tightenlines,nofootinbib]{revtex4}
\usepackage{graphicx}
\usepackage{epsfig}
\usepackage{dcolumn}
\usepackage{bm}
\usepackage{overpic}
\usepackage{diagbox}
\usepackage{color} 
\begin{document}

\normalsize
\parskip=0pt plus 1pt minus 1pt

\title{\boldmath Observation of $e^{+}e^{-} \to \eta h_{c}$ at center-of-mass energies from 4.085 to 4.600\,GeV}

\author{
M.~Ablikim$^{1}$, M.~N.~Achasov$^{9,d}$, S. ~Ahmed$^{14}$, X.~C.~Ai$^{1}$, O.~Albayrak$^{5}$, M.~Albrecht$^{4}$, D.~J.~Ambrose$^{45}$, A.~Amoroso$^{50A,50C}$, F.~F.~An$^{1}$, Q.~An$^{47,38}$
% ...
\\
\vspace{0.2cm}
(BESIII Collaboration)\\
\vspace{0.2cm} {\it
$^{1}$ Institute of High Energy Physics, Beijing 100049, People's Republic of China\\
% ...
$^{4}$ Bochum Ruhr-University, D-44780 Bochum, Germany\\
$^{5}$ Carnegie Mellon University, Pittsburgh, Pennsylvania 15213, USA\\
% ...
$^{9}$ G.I. Budker Institute of Nuclear Physics SB RAS (BINP), Novosibirsk 630090, Russia\\
% ...
$^{14}$ Helmholtz Institute Mainz, Johann-Joachim-Becher-Weg 45, D-55099 Mainz, Germany\\
% ...
$^{38}$ State Key Laboratory of Particle Detection and Electronics, Beijing 100049, Hefei 230026, People's Republic of China\\
% ...
$^{45}$ University of Rochester, Rochester, New York 14627, USA\\
% ...
$^{47}$ University of Science and Technology of China, Hefei 230026, People's Republic of China\\
% ...
$^{50}$ (A)University of Turin, I-10125, Turin, Italy; (B)University of Eastern Piedmont, I-15121, Alessandria, Italy; (C)INFN, I-10125, Turin, Italy\\
% ...
$^{d}$ Also at the Novosibirsk State University, Novosibirsk, 630090, Russia\\
% ...
}
}

\date{\today}

\begin{abstract}

We observe for the first time the process $e^{+}e^{-} \rightarrow \eta
h_c$ with data collected by the BESIII experiment. Significant signals
are observed at the center-of-mass energy $\sqrt{s}=4.226$\,GeV, and
the Born cross section is measured to be $(9.5^{+2.2}_{-2.0} \pm
2.7)$\,pb. Evidence for $\eta h_c$ is observed at
$\sqrt{s}=4.358$\,GeV with a Born cross section of
$(10.0^{+3.1}_{-2.7} \pm 2.6)$\,pb, and upper limits on the production
cross section at other center-of-mass energies between 4.085 and
4.600\,GeV are determined.
\end{abstract}

\maketitle

\section{Introduction}

The spectroscopy of charmonium states below the open charm threshold is well established, but the
situation above the threshold is more complicated. From the inclusive hadronic cross section in
$e^+e^-$ annihilation, some vector charmonium states, $\psi(3770)$, $\psi(4040)$, $\psi(4160)$,
$\psi(4415)$ are known with properties as expected in the quark
model~\cite{Ablikim:2007gd}. However, besides these states, several new vector states, namely the
Y(4260), Y(4360) and Y(4660), have been discovered experimentally~\cite{Aubert:2005rm,He:2006kg,Yuan:2007sj,Ablikim:2015tbp,Aubert:2007zz, Wang:2007ea}. 
In addition, some new states with other
quantum number configurations are also found in experiment, such as the $X(3872)$, $Z_{c}(3900)$ and
$Z_{c}(4020)$ states~\cite{Choi:2003ue,Ablikim:2013mio,Liu:2013dau,Ablikim:2013xfr,Ablikim:2015tbp,Ablikim:2015gda,Ablikim:2013wzq,Ablikim:2013emm,Ablikim:2014dxl,Ablikim:2015vvn}. The
common properties of these states are their relatively narrow width for decaying into a pair of
charmed mesons, and their strong coupling to hidden charm final states. Therefore, it is hard to
explain all these resonances as charmonia and they are named `charmonium-like states' collectively.
Several unconventional explanations, such as hybrid charmonium~\cite{Close:2005iz,Zhu:2005hp,Chen:2016ejo},
tetraquark~\cite{Ebert:2005nc,Maiani:2005pe,TWChiu}, hadronic
molecule~\cite{Liu:2005ay,CFQiao,Yuan:2005dr}, diquarks~\cite{Chen:2015dig,Padmanath:2015era} or kinematical effects~\cite{Bugg:2008wu,Chen:2011xk,Wang:2013cya,Swanson:2014tra} have been suggested.
See also Ref.~\cite{Wang:2013kra,Lebed:2016hpi} and references therein for a recent review.

To understand the nature of these charmonium-like states, it is mandatory to investigate both open and hidden charm
decays. Most of the observed vector charmonium-like states transit to spin-triplet charmonium states with large rate
since the spin alignment of the $c$ and $\bar{c}$-quarks does not need to be changed between initial and final
states. However, the spin-flip process $e^{+}e^{-}\to\pi\pi h_c$ has also been observed by the CLEO~\cite{CLEO:2011aa} and BESIII
experiments~\cite{Ablikim:2013wzq,Ablikim:2014dxl,BESIII:2016adj}, and the large cross section exceeds theoretical
expectations~\cite{Voloshin:2004mh}. Furthermore, two new structures have been reported in
$e^+e^- \to \pi^+\pi^- h_c$~\cite{BESIII:2016adj}.  This may suggest the existence of hybrid charmonium states with a
pair of $c\bar{c}$ in spin-singlet configuration which easily couples to an $h_c$ final state.  
Consequently, searching for the process $e^{+}e^{-}\to\eta h_c$ will provide more information about the spin-flip transition, and the structures observed in $e^{+}e^{-}\to\pi\pi h_c$ may be observed  also in the $\eta h_c$ process. In addition, the transition $\Upsilon(4S)\to\eta h_b$ has been observed in the bottomonium system~\cite{Tamponi:2015xzb}. The analogous process in the
charmonium system is worth searching for to understand the dynamics in the $\eta$ transition between heavy quarkonia. 

The CLEO Collaboration observed evidence of about $3\sigma$ for $e^{+}e^{-}\to \eta h_c $ based on
$586\,\rm{pb^{-1}}$ data taken at $\sqrt{s}=4.17$\,GeV~\cite{CLEO:2011aa}, and the measured
cross section is $ (4.7\pm 2.2)\,\rm{pb}$. In comparison, BESIII has collected data samples of about
$4.7\,\rm{fb^{-1}}$ in total at $\sqrt{s}>4.0$\,GeV.  In this paper, a search is performed for
the process $e^{+}e^{-}\to\eta h_c$ with $h_c \to \gamma \eta_c$ based on data samples collected with the
BESIII detector at center-of-mass (c.m.) energies from 4.085 to 4.600\,GeV, as listed in
Table~\ref{Finalresults}. The integrated luminosities of these data samples are measured by
analyzing large-angle Bhabha scattering events with an uncertainty of
$1.0\%$~\cite{Ablikim:2015nan}, and the c.m.\ energies are measured using the di-muon
process~\cite{Ablikim:2015zaa}. In the analysis, $\eta_c$ is reconstructed with 16 hadronic final states:
$p\bar{p}$, $2(\pi^+ \pi^-)$, $2(K^+ K^-)$, $K^+ K^- \pi^+ \pi^-$, $p\bar{p} \pi^+ \pi^-$,
$3(\pi^+ \pi^-)$, $K^+ K^- 2(\pi^+ \pi^-)$, $K^+ K^- \pi^0$, $p \bar{p}\pi^0$,
$K^0_S K^\pm \pi^\mp$, $K^0_S K^\pm\pi^\mp \pi^\pm \pi^\mp$, $\pi^+ \pi^- \eta$, $K^+ K^- \eta$,
$2(\pi^+ \pi^-) \eta$, $\pi^+ \pi^- \pi^0 \pi^0$, and $2(\pi^+\pi^-) \pi^0 \pi^0$, in which $K_S^0$
is reconstructed from its $\pi^+\pi^-$ decay, and $\pi^0$ and $\eta$ from their $\gamma \gamma$ final
state.

\section{Detector and data samples}

BEPCII is a two-ring $e^+e^-$ collider designed for a peak luminosity of $10^{33}$\,cm$^{-2}s^{-1}$
at a beam current of 0.93\,A per beam. The cylindrical core of the BESIII detector consists of a
helium-gas-based main drift chamber (MDC) for charged-particle tracking and particle identification
(PID) through the specific energy loss d$E$/d$x$, a plastic scintillator time-of-flight (TOF) system 
for additional PID, and
a 6240-crystal CsI(Tl) electromagnetic calorimeter~(EMC) for electron identification and photon
detection. These components are all enclosed in a superconducting solenoidal magnet providing a 1-T
magnetic field.  The solenoid is supported by an octagonal flux-return yoke instrumented with
resistive-plate-counter muon detector modules interleaved with steel.  The geometrical acceptance
for charged tracks and photons is $93\%$ of $4\pi$, and the resolutions for charged-track momentum
at 1\,GeV is $0.5\%$. The resolutions of photon energy in barrel and end-cap regions are $2.5\%$ and $5\%$, respectively.  More details on the features and
capabilities of BESIII are provided in Ref.~\cite{ref:bes3}.

A Monte Carlo (MC) simulation is used to determine the detection efficiency and to estimate physics background. The detector response is modelled with a {\sc geant4}-based~\cite{Agostinelli:2002hh,Allison:2006ve} detector simulation package.
Signal and background processes are generated with specialized models that have been packaged and customized for BESIII. 40,000 MC
events are generated for each decay mode of $\eta_c$ at each c.m.\ energy with {\sc
  kkmc}~\cite{ref:kkmc} and {\sc besevtgen}~\cite{ref:bes3gen}. The events are generated with an
$h_c$ mass of $3525.28$\,MeV/$c^2$ and a width of $1.0$\,MeV.  The $E$1 transition
$h_c\to\gamma\eta_c$ is generated with an angular distribution of $1 +\cos^2\theta^*$, where
$\theta^*$ is the angle of the $E$1 photon with respect to the $h_c$ helicity direction in the $h_c$
rest frame. Multi-body $\eta_c$ decays are generated uniformly in phase space.  
In order to study potential backgrounds, inclusive MC samples with the same size as the data are produced at $\sqrt{s}=4.23$, 4.26 and 4.36\,GeV. They are generated using {\sc kkmc}, which includes the decay of
$Y(4260)$, ISR production of the vector charmonium states, charmed meson production, QED events, and
continuum processes. The known decay modes of the resonances are generated with {\sc besevtgen} with
branching fractions set to the world average values~\cite{Olive:2016xmw}.  The remaining charmonium
decays are generated with {\sc lundcharm}~\cite{Chen:2000tv}, while other hadronic events are
generated with {\sc pythia}~\cite{Sjostrand:2001yu}.

\section{Event Selection and Study of Background}

According to the MC simulation of $e^{+}e^{-}\to\eta h_c$ with $h_c\to\gamma\eta_c$ at
$\sqrt{s}=4.226$\,GeV, the energy of the photon emitted in the $E$1 transition $h_c \to \gamma \eta_c$
is expected to be in the range (400, 600)\,MeV in the laboratory frame. Therefore, the signal event should have
one $E$1 photon candidate with energy located in the expected region and one $\eta$ candidate with
recoil mass in the region of $(3480,3600)\,\rm{MeV}/c^2$. We define the $\eta$ recoil mass
$M_\text{recoil}(\eta)$ as $M_\text{recoil}(\eta)^{2}c^{4}\equiv(E_\text{cm}-E_{\eta})^{2}-|\vec p_\text{cm}-\vec p_{\eta}|^{2}c^{2}$,
where $(E_\text{cm},\vec p_\text{cm})$ and $(E_{\eta},\vec p_{\eta})$ are the four-momenta of the $e^{+}e^{-}$ system
and $\eta$ in the $e^{+}e^{-}$ rest frame.  Since the $E$1 photon energy distribution in the laboratory frame will broaden
with increasing c.m.\ energy, the energy window requirement is enlarged to (350, 650)\,MeV
for the data sets collected at $\sqrt{s}>4.416$\,GeV. The $\eta_c$ candidate is reconstructed by the
hadronic systems determined by the corresponding decay mode. The invariant mass of the hadronic systems
is required to be within the mass range of $(2940, 3020)\,\rm{MeV}/c^2$. For the selected
candidates, we apply a fit to the distribution of the $\eta$ recoil mass to obtain the signal yield.

Charged tracks in BESIII are reconstructed from MDC hits within a fiducial range of
$|\cos\theta| < 0.93$, where $\theta$ is the polar angle of the track.  We require that the point of closest
approach (POCA) to the interaction point (IP) is within 10\,cm in the beam direction and within 1\,cm
in the plane perpendicular to the beam direction. A vertex fit constrains the production vertex,
which is determined run by run, and all the charged tracks to a common vertex.  Since the $K_{S}^{0}$
has a relatively long lifetime, it will travel a certain distance in the detector to the point where it decays into
daughter particles.  The requirements on the track POCA and the vertex fit mentioned above are therefore not
applied to its daughter particles.  The TOF and d$E/$d$x$ information are combined to form PID
confidence levels (C.L.) for the pion, kaon, and proton hypotheses; both PID and kinematic fit
information is used to determine the particle type of each charged track, as discussed below.

Electromagnetic showers are reconstructed by clustering EMC crystal energies. Efficiency and energy
resolution are improved by including energy deposits in nearby TOF counters. A photon candidate is
defined by showers detected with the EMC exceeding a threshold of 25\,MeV in the barrel region
($|\cos \theta| < 0.8$) or of 50\,MeV in the end-cap region ($0.86 < |\cos\theta| < 0.92$). Showers
in the transition region between the barrel and the end-cap are excluded because of the poor
reconstruction. Moreover, EMC cluster timing requirements are used to suppress electronic noise and
energy deposits unrelated to the event.

Candidates for $\pi^0$ ($\eta$) mesons are reconstructed from pairs of photons with an invariant mass
$M(\gamma\gamma)$ satisfying $|M(\gamma\gamma)-m_{\pi^0(\eta)}|<15\,\rm{MeV}/c^2$. A one-constraint (1C) kinematic fit with the
$M(\gamma\gamma)$ constrained to the $\pi^0$ ($\eta$) nominal mass
$m_{\pi^{0}}$ ($m_{\eta}$)~\cite{Olive:2016xmw} is performed to improve the energy resolution. We
reconstruct $K^0_S\to\pi^+\pi^-$ candidates with pairs of oppositely charged tracks with an
invariant mass in the mass range of $|M(\pi\pi)-m_{K_S}| < 20\,\rm{MeV}/c^2$. Here, $m_{K_S}$
denotes the nominal mass of $K_{S}^{0}$~\cite{Olive:2016xmw}. A vertex fit constrains the charged
tracks to a common decay vertex, and the corrected track parameters are used to calculate the
invariant mass. To reject random $\pi^+\pi^-$ combinations, a kinematic constraint between the
production and decay vertices, called a secondary-vertex fit, is
employed~\cite{ref::ks0-reconstruction}, and the decay length is required to be more than twice the
vertex resolution.

The $\eta_c$ candidate is reconstructed in its decay to one of the 16 decay modes mentioned earlier. After the above
selection, a four-constraint (4C) kinematic fit is performed for each event imposing overall
energy-momentum conservation, and the $\chi^2_{\rm 4C}$ is required to be less than 25 to suppress
background events with different final states. If multiple $\eta_c$ candidates are found in an
event, only the one with the smallest
$\chi^2 \equiv \chi^2_{\rm 4C}+\chi^2_{\rm 1C}+\chi^2_{\rm pid}+\chi^2_{\rm vertex}$ is retained, where
$\chi^2_{\rm 1C}$ is the $\chi^2$ of the 1C fit for $\pi^0$ ($\eta$), $\chi^2_{\rm pid}$ is the sum over all charged tracks of the
$\chi^2$ of the PID hypotheses, and $\chi^2_{\rm vertex}$ is the $\chi^2$ of the
$K_S^0$ secondary-vertex fit. If more than one $\eta$ candidate with recoil mass in the $h_c$ signal region
($3480<M_\text{recoil}(\eta)<3600\,\rm{MeV}/c^2$) is found, the one which leads to a mass of the
$\eta_c$ candidate closest to the $\eta_{c}$ nominal mass $m_{\eta_c}$ is selected to reconstruct
the $\eta_c$.

The requirement on $\chi_{4C}^{2}$ and mass (energy) windows for $\eta$, $\eta_{c}$ and $E$1 photon 
reconstruction are determined by maximizing the figure-of-merit, 
${\rm FOM}=N_{S}/\sqrt{N_{S}+N_{B}}$, where $N_{S}$ represents the number of signal events
determined by MC simulation, and $N_{B}$ represents the number of background events
obtained from $h_c$ sidebands in the data sample. The cross section of $e^+e^-\to \eta h_c$ measured
by CLEO~\cite{CLEO:2011aa} and the $\eta_c$ branching ratios given by the Particle Data Group
(PDG)~\cite{Olive:2016xmw} are used to scale the number of signal events in the optimization.

\begin{figure*}[htbp]
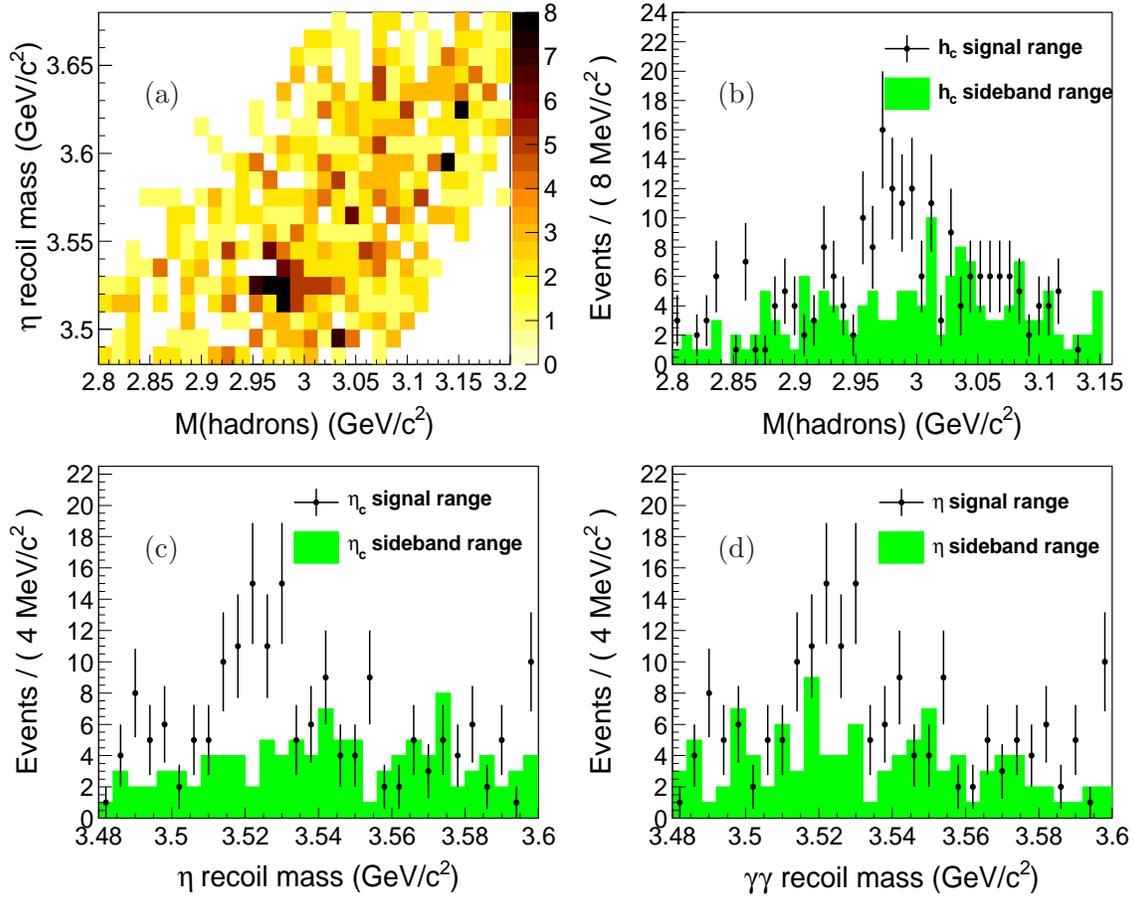

\begin{center}
\begin{overpic}[width=7.5cm,height=6cm,angle=0]{Fig1a_etahc_etac_hc_all_4230.eps}
\put(25,60){(a)}
\end{overpic}
\begin{overpic}[width=7.5cm,height=6cm,angle=0]{Fig1b_etahc_etac_sig_sid_sum_4230.eps}
\put(25,60){(b)}
\end{overpic}

\begin{overpic}[width=7.5cm,height=6cm,angle=0]{Fig1c_etahc_hc_sig_sid_sum_4230.eps}
\put(25,60){(c)}
\end{overpic}
\begin{overpic}[width=7.5cm,height=6cm,angle=0]{Fig1d_etahc_hc_eta_sig_sid_sum_4230.eps}
\put(25,60){(d)}
\end{overpic}
\end{center}
\caption{Mass spectrum obtained at $\sqrt{s}=4.226$\,GeV. (a) The two-dimensional distribution of the
  invariant mass of the hadronic system and the recoil mass of $\eta$; (b) mass of hadrons in $h_{c}$ signal ([3.51,
  3.55]\,GeV$/c^{2}$) and sideband regions ([3.48, 3.50]\,GeV$/c^{2}$ and [3.56, 3.58]\,GeV$/c^{2}$); (c) $\eta$ recoil mass
  in $\eta_c$ signal ([2.94, 3.02]\,GeV$/c^{2}$) and sideband region ([2.87, 2.91]\,GeV$/c^{2}$ and
  [3.05,3.09]\,GeV$/c^{2}$), and (d) $\gamma\gamma$ recoil mass in $\eta$ signal ([0.531,0.563]\,GeV$/c^{2}$) and
  sideband regions ([0.505, 0.521]\,GeV$/c^{2}$ and [0.573, 0.589]\,GeV$/c^{2}$). For(b), (c), and (d), the dots with
  error bars represent the distributions in the signal regions and the shaded histograms represent the
  distributions in the sidebands.}\label{hc_sig_sid}
\end{figure*}

After applying all the criteria to the data sample taken at $\sqrt{s}=$4.226\,GeV, the events
cluster in the signal region in the two-dimensional distribution as shown in
Fig.~\ref{hc_sig_sid}(a). If the two-dimensional histogram is projected to each axis, clear $\eta_c$
and $h_c$ signals can be found in the expected regions as shown in Fig.~\ref{hc_sig_sid}(b) and
(c). Meanwhile, no structure is observed in the events from the $\eta_c$ ($h_c$) sideband regions.  To
further understand the background shape, events located in the $\eta$ sideband regions are also
investigated, which are shown by the green shaded area in Fig.~\ref{hc_sig_sid} (d) and are
well described by a smooth distribution.

In addition, inclusive MC samples generated at $\sqrt{s}=4.23$\,GeV are analyzed to study the
background components. Here, the ratios among different components are fixed according to
theoretical calculation or experimental measurements, except for the Bhabha process. A sample of 
$1.0\times10^{7}$ Bhabha events (about $2\%$ of the Bhabha events in real data) 
is generated with the {\sc Babayaga} generator~\cite{Balossini:2006wc}
for background estimation. From this study, the dominant background sources are found to be
continuum processes according to the MC truth information, while $Y(4260)$ decays only give a small
contribution to the total background. Most background events from resonance decays are
$\pi\pi J/\psi$, $\omega \chi_{c0}$ and open charm production. A similar conclusion can be drawn for
data samples taken at other c.m.\ energies. From the study above, we conclude that the background
shape in the $\eta$ recoil mass can be described by a linear function.

\section{\boldmath Fit to the recoil mass of $\eta$}

\begin{figure*}[tbhp]
\centering
\includegraphics[width=16cm]{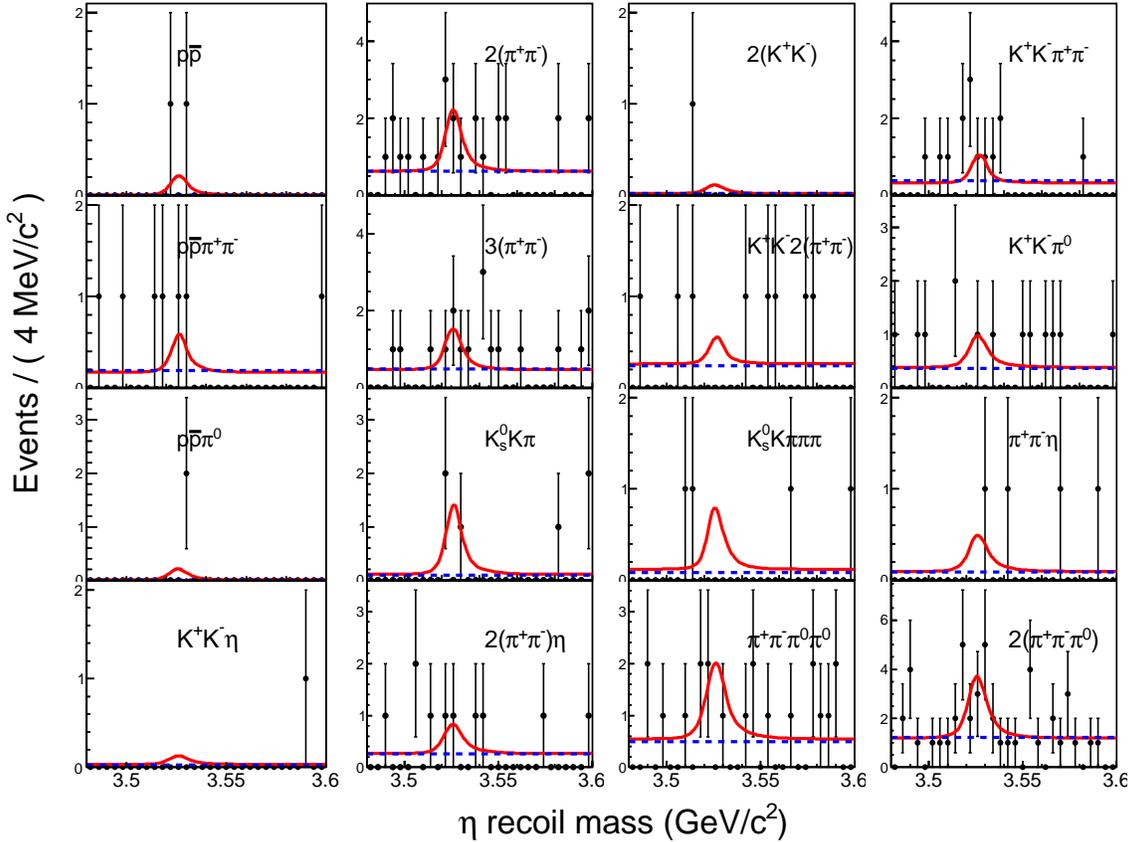}
\caption{Simultaneously fitted $\eta$ recoil mass spectra in
  $e^+e^-\to\eta{h}_c, {h}_c\to\gamma\eta_c$, $\eta_c \to X_i$ for the 16 final states $X_i$ at
  $\sqrt{s}=$4.226\,GeV. The dots with error bars represent the $\eta$ recoil mass spectrum in
  data. The solid lines show the total fit function and the dashed lines are the background
  component of the fit. }
\label{fig:fitRMetasim}
\end{figure*}

\begin{figure}[tbhp]
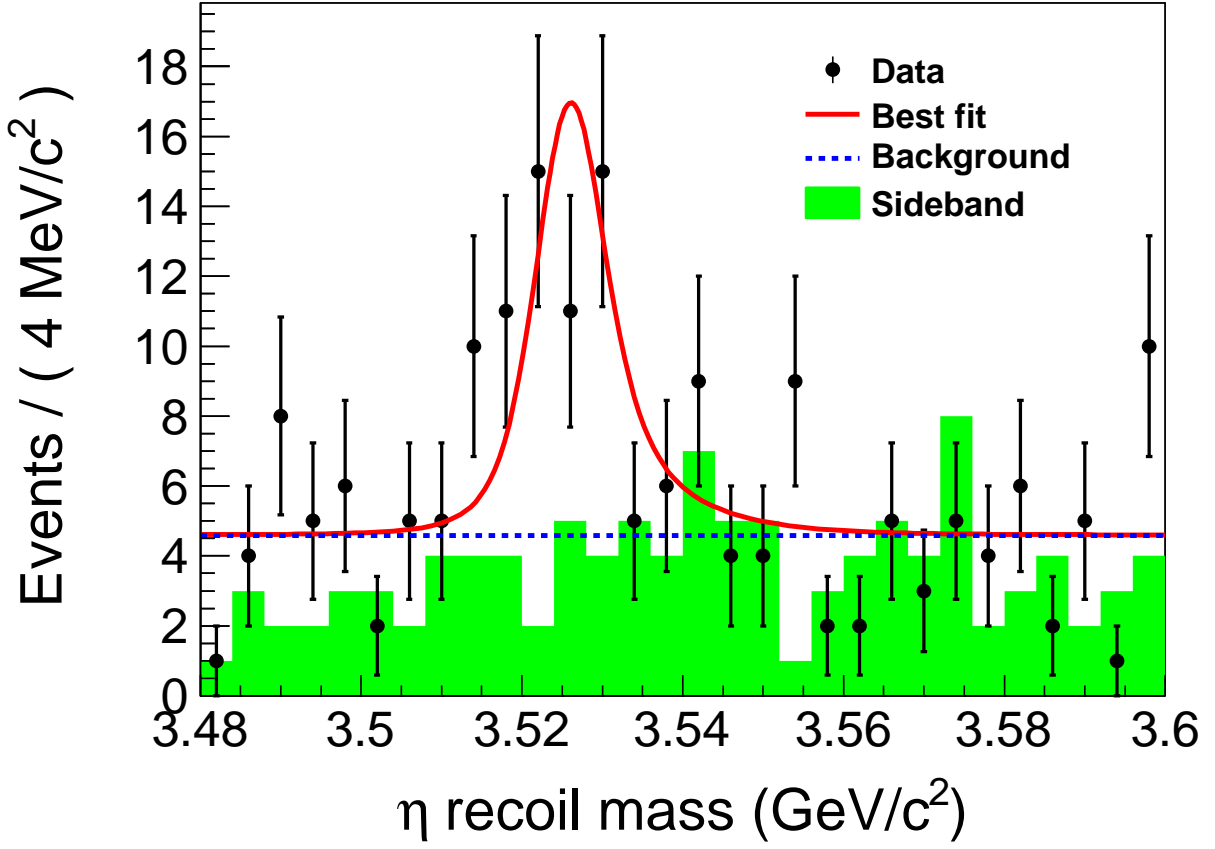

\begin{center}
\begin{overpic}[width=\columnwidth,angle=0]{Fig3_simulfit_psi_4230_eta_hc_16ch_sum.eps}
\end{overpic}
\caption{Sum of the simultaneous fits to $\eta$ recoil mass spectra for all 16 $\eta_c$ decay modes
  at $\sqrt{s}=4.226$\,GeV. The dots with error bars represent the $\eta$ recoil mass spectrum in
  data. The solid line shows the total fit function and the dashed line is the  background component
  of the fit. The shaded histogram shows the events from the $\eta_c$ sidebands.
  \label{fig:fitRMetasimtot}}
\end{center}
\end{figure}

To obtain the $h_c$ yield for each $\eta_c$ decay channel, the 16 $\eta$ recoil mass distributions
are fitted simultaneously using an unbinned maximum likelihood method.  In the fit, the signal shape
is determined by the MC simulation and the background shape is described by a linear function. The total
signal yield of 16 channels is set to be $N_{\rm obs}$, which is the common variable for all
sub-samples and required to be positive. $N_{\rm obs}\times f_i$ is the signal yield of the $i$-th
channel. Here, $f_i$ refers to the weight factor
$f_i \equiv \mathcal{B}_{i} \epsilon_{i}/\sum \epsilon_{i} \mathcal{B}_i$, in which the $\mathcal{B}_{i}$
denotes the branching fraction of $\eta_c$ decays to the $i$-th final state and $\epsilon_{i}$
represents the corresponding efficiency. The efficiency for two-body $\eta_{c}$ decays is about $20\%$, for three- or four-body decays is about $10\%$ and for six-body decays it is about $6\%$.  The signal
and the background normalization for each mode are free parameters in the fit. The mode-by-mode and summed fit
results are shown in Figs.~\ref{fig:fitRMetasim} and \ref{fig:fitRMetasimtot}, respectively. The
$\chi^2$ per degree of freedom (dof) for this fit is $\chi^2/\rm{dof}=17.2/15=1.15$, where sparsely
populated bins are combined so that there are at least 7 counts per bin in the $\chi^2$
calculation. The total signal yield is $41\pm9$ with a statistical significance of $5.8\,\sigma$.

\begin{figure}[tbhp]
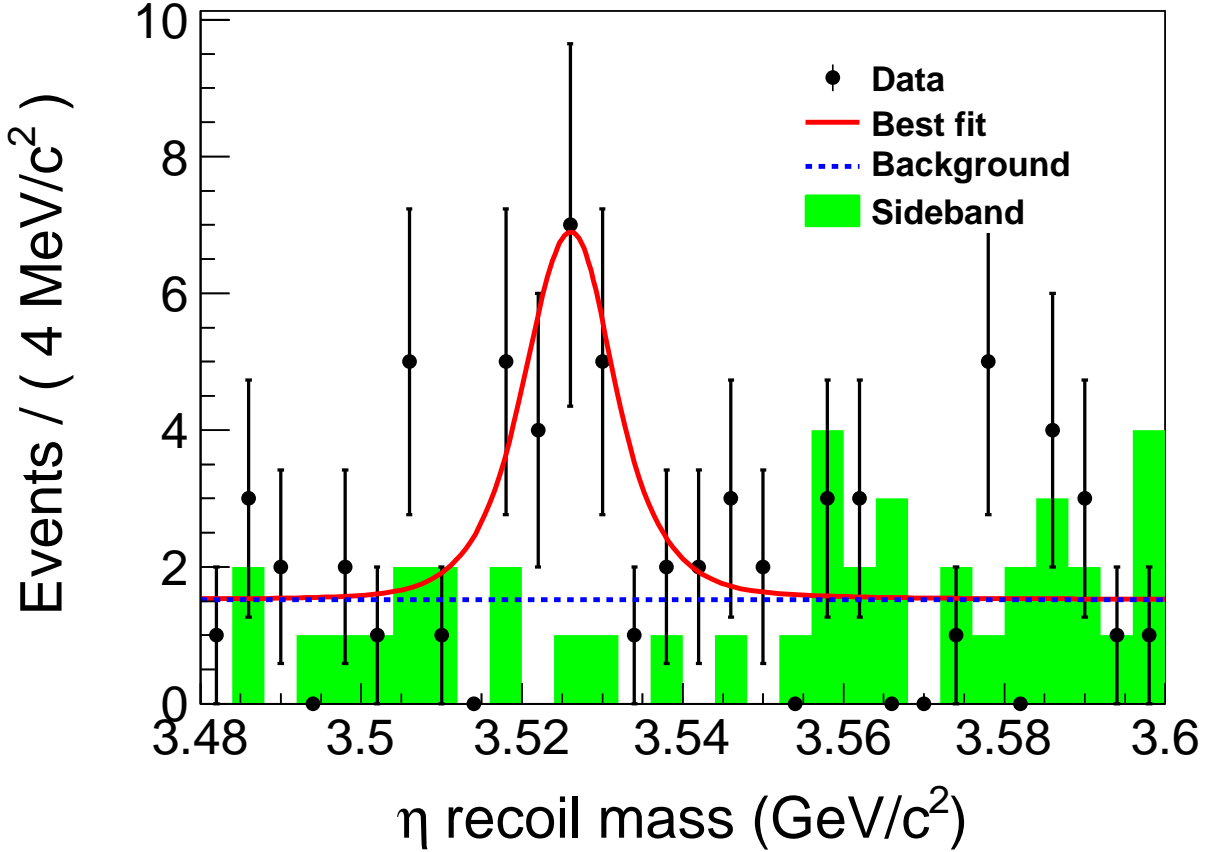

\begin{center}
\begin{overpic}[width=\columnwidth,angle=0]{Fig4_simulfit_psi_4360_eta_hc_16ch_sum.eps}
\end{overpic}
\caption{Sum of the simultaneous fits to $\eta$ recoil mass spectra for all 16 $\eta_c$ decay modes
  at $\sqrt{s}=4.358$\,GeV. The dots with error bars represent the $\eta$ recoil mass spectrum in
  data. The solid line shows the total fit function and the dashed line is the  background component
  of the fit. The shaded histogram shows the events from $\eta_c$ sidebands.
  \label{fig:fithcpi0tot_4360}}
\end{center}
\end{figure}

With the same method, evidence for $e^+e^-\to\eta h_c$ is found in the data sample taken at
$\sqrt{s}=4.358$\,GeV, as shown in Fig.~\ref{fig:fithcpi0tot_4360}, but no obvious signals are
observed for the data sets taken at other c.m.\ energies.

\begin{widetext}
\section{Born cross section measurement}
The Born cross section is calculated using the following formula:

\begin{equation}\label{eq1}
\sigma^{\rm Born}(e^+e^-\to \eta h_c)=\frac{N_{\rm obs}}{\mathcal{L} (1+\delta) |1+\Pi|^2 \mathcal{B}(\eta\to\gamma \gamma) \mathcal{B}({h}_c\to\gamma \eta_c)  \Sigma_{i}\epsilon_{i} \mathcal{B}_i}.
\end{equation}
\end{widetext}

Here, $\mathcal{L}$ is the integrated luminosity of the data sample taken at each c.m.\
energy. $(1+\delta)$ is the radiative correction factor, which is defined as
\begin{equation}\label{eq2}
(1+\delta) = \frac{ \int \sigma(s(1-x)) F(x,s)dx}{\sigma(s)}, 
\end{equation}
where $F(x,s)$ is the radiator function, which is known from a QED calculation with an accuracy of
$0.1\%$~\cite{Kuraev:1985hb}. Here, $s$ is squared c.m. energy, and $s(1-x)$ is the squared c.m. energy after emission of the ISR photons. $\sigma(s)$ is the energy dependent Born cross section in the
range of [4.07, 4.6]\,GeV. Actually, the radiative correction depends on the Born cross section from
the production threshold to the $e^+e^-$ collision energy, which is also what we want to measure in
this analysis. Therefore, the final Born cross section is obtained in an iterative way.  The efficiencies
from a set of signal MC samples without any radiative correction are used to calculate a first approximation to the observed
cross section. Then, by taking the observed cross sections as inputs, new MC samples are
generated with radiative correction and the efficiencies as well as $(1+\delta)$ are updated. After
that, the cross sections can also be recalculated accordingly. The iterations are performed in this
way until a stable result is obtained. The values of $(1+\delta)$ from the last iteration are shown
in Table~\ref{Finalresults}.

The term $|1+\Pi|^2$ is the vacuum-polarization (VP) correction factor, which includes leptonic and
hadronic contributions. This factor is calculated with the package provided in
Ref.~\cite{Eidelman:1995ny}. The package provides leptonic and hadronic VP both in the space-like
and time-like regions. For the leptonic VP the complete one- and two-loop results and the known
high-energy approximation for the three-loop corrections are included. The hadronic contributions
are given in tabulated form in the subroutine {\sc hard5n}~\cite{Jegerlehner:2011mw}. The $|1+\Pi|^2$ values are also shown in
Table~\ref{Finalresults}.

\begin{figure}[htbp]
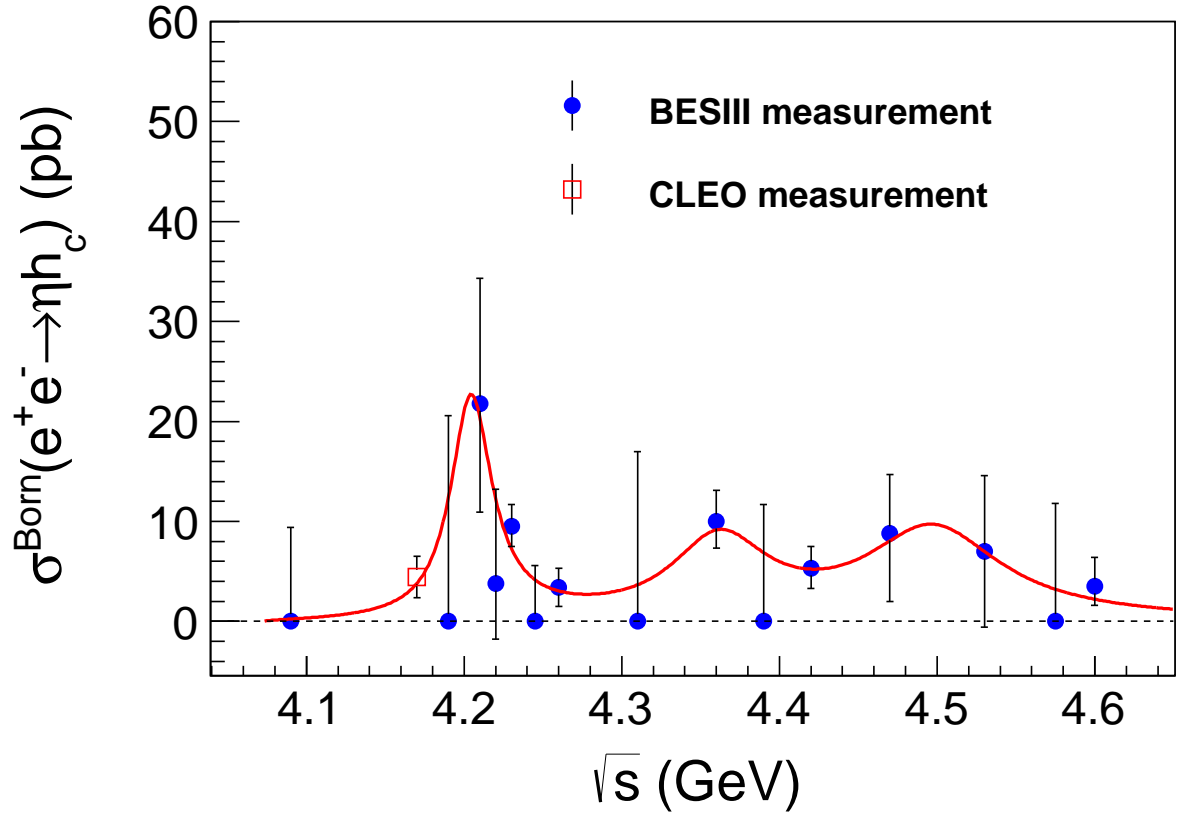

\begin{center}
\begin{overpic}[width=\columnwidth,angle=0]{Fig5_Fit_cross_section_vs_ecm_17_points_Ase.eps}
\end{overpic}
\end{center}
\caption{Fit to the cross section of $e^+e^-\to\eta{h}_c$ as a function of c.m.\ energies. The square
  with error bar shows the measurement from CLEO~\cite{CLEO:2011aa}, the dots with error bars
  refer to the results of this measurement, and the solid line shows the fit result with 3 coherent
  BW functions.  }\label{FitBESandBELLE}
\end{figure}

Table~\ref{Finalresults} and Fig.~\ref{FitBESandBELLE} show the energy dependent Born cross sections
from this measurement. Taking into account the CLEO measurement at
$\sqrt{s}=4.17$\,GeV~\cite{CLEO:2011aa}, the cross section from $4.085\sim4.600$\,GeV is
parameterized as the coherent sum of three Breit-Wigner (BW) functions, as shown by the solid line in
Fig.~\ref{FitBESandBELLE}.  In the fit, the parameters of the BW around 4.36\,GeV are fixed to
those of the $Y(4360)$~\cite{Wang:2007ea} while the parameters of the other two BW functions are left free in the fit. The fitted
parameters of the free BW are: $\rm{M_{1}} = (4204\pm6$)\,\rm MeV/$c^2$,
$\Gamma_{1} = (32 \pm 22)$\,MeV and $\rm{M_{2}} = (4496\pm26$)\,\rm MeV/$c^2$,
$\Gamma_{2} = (104 \pm 69)$\,MeV, where the uncertainties are statistical.

\begin{table*}[htbp]
  \centering
   \caption{ Data sets and results of the Born cross section measurement for $e^{+}e^{-} \rightarrow
     \eta h_c$. The table includes the integrated luminosity $\mathcal{L}$, the number of observed
     signal events $N_{\rm obs}$, the radiative correction $(1+\delta)$ and vacuum polarization
     correction factor $|1+\Pi|^2$, the sum of the products of the branching fraction and
     efficiency $\sum \epsilon_{i} \mathcal{B}_i$, the Born cross section $\sigma^B$ and its upper
     limit (at the 90\% C.L.), and the statistical significance $\mathcal{S}$.}
  \begin{tabular*}{\textwidth}{cccccccc}
  \hline
  \hline
   $\sqrt{s}$\,(MeV)  &    $\mathcal{L}$\,($\rm{pb^{-1}}$)   &  $N_{\rm{obs}}$  & $(1+\delta)$  & $|1+\Pi|^2$ & $\sum \epsilon_{i} \mathcal{B}_i$ ($10^{-2}$)  & $\sigma^B$\,($\rm{pb}$)  & \hspace{2em}  $\mathcal{S}$  \\
  \hline
4085.4 & 52.4 & $ 0.0^{+1.7}_{-0} $ & 0.68 & 1.052 & 2.40 & $ 0.0^{+9.4}_{-0} \pm 5.4 $ ($< 23.7 $) & \hspace{2em} 0.0\,$\sigma$ \\
4188.6 & 43.1 & $ 0.0^{+2.9}_{-0} $ & 0.69 & 1.056 & 2.24 & $ 0.0^{+20.6}_{-0} \pm 13.7 $ ($< 52.2 $) & \hspace{2em} 0.0\,$\sigma$ \\
4207.7 & 54.6 & $ 4.2^{+2.4}_{-2.1} $ & 0.75 & 1.057 & 2.22 & $ 21.8^{+12.5}_{-10.9} \pm 5.7 $ ($< 53.6 $) & \hspace{2em} 1.7\,$\sigma$ \\
4217.1 & 54.1 & $ 0.8^{+2.0}_{-1.2} $ & 0.85 & 1.057 & 2.18 & $ 3.8^{+9.4}_{-5.6} \pm 1.0 $ ($< 32.2 $) & \hspace{2em} 0.5\,$\sigma$ \\
4226.3 & 1091.7 & $ 41.2^{+9.5}_{-8.7} $ & 0.95 & 1.056 & 1.97 & $ 9.5^{+2.2}_{-2.0} \pm 2.7 $ & \hspace{2em} 5.8\,$\sigma$ \\
4241.7 & 55.6 & $ 0.0^{+1.2}_{-0} $ & 1.06 & 1.056 & 1.72 & $ 0.0^{+5.6}_{-0} \pm 5.0 $ ($< 17.6 $) & \hspace{2em} 0.0\,$\sigma$ \\
4258.0 & 825.7 & $ 10.3^{+5.8}_{-5.6} $ & 1.11 & 1.054 & 1.56 & $ 3.4^{+1.9}_{-1.9} \pm 1.2 $ ($< 8.3 $) & \hspace{2em} 2.0\,$\sigma$ \\
4307.9 & 44.9 & $ 0.0^{+2.7}_{-0} $ & 0.93 & 1.052 & 1.80 & $ 0.0^{+17.0}_{-0} \pm 8.4 $ ($< 35.3 $) & \hspace{2em} 0.0\,$\sigma$ \\
4358.3 & 539.8 & $ 19.0^{+5.9}_{-5.2} $ & 0.81 & 1.051 & 2.07 & $ 10.0^{+3.1}_{-2.7} \pm 2.6 $ ($< 19.3 $) & \hspace{2em} 4.3\,$\sigma$ \\
4387.4 & 55.2 & $ 0.0^{+2.3}_{-0} $ & 0.90 & 1.051 & 1.87 & $ 0.0^{+11.7}_{-0} \pm 5.8 $ ($< 26.2 $) & \hspace{2em} 0.0\,$\sigma$ \\
4415.6 & 1073.6 & $ 18.6^{+7.8}_{-7.2} $ & 0.94 & 1.053 & 1.65 & $ 5.3^{+2.2}_{-2.0} \pm 1.4 $ ($< 11.2 $) & \hspace{2em} 2.9\,$\sigma$ \\
4467.1 & 109.9 & $ 3.1^{+2.1}_{-2.4} $ & 0.85 & 1.055 & 1.79 & $ 8.8^{+5.9}_{-6.8} \pm 2.3 $ ($< 19.0 $) & \hspace{2em} 1.1\,$\sigma$ \\
4527.1 & 110.0 & $ 2.1^{+2.3}_{-2.3} $ & 0.94 & 1.055 & 1.38 & $ 7.0^{+7.6}_{-7.6} \pm 1.8 $ ($< 27.7 $) & \hspace{2em} 0.8\,$\sigma$ \\
4574.5 & 47.7 & $ 0.0^{+1.2}_{-0} $ & 1.15 & 1.055 & 0.88 & $ 0.0^{+11.8}_{-0} \pm 6.8 $ ($< 28.6 $) & \hspace{2em} 0.0\,$\sigma$ \\
4599.5 & 566.9 & $ 4.0^{+3.3}_{-2.2} $ & 1.27 & 1.055 & 0.75 & $ 3.5^{+2.9}_{-1.9} \pm 0.9 $ ($< 11.1 $) & \hspace{2em} 1.7\,$\sigma$ \\
\hline
  \hline
    \end{tabular*}
    \label{Finalresults}
\end{table*}

\section{Systematic uncertainties}

In this section, the study of the systematic uncertainty for the cross section measurement at
$\sqrt{s}=4.226$\,GeV is described.  The same method is applied to the other c.m.\ energies.

The main contributions to the systematic uncertainties are from the luminosity measurement, the fit
method, $\mathcal{B}({h}_c\to\gamma \eta_c)\mathcal{B}(\eta\to\gamma \gamma)$, ISR correction, VP correction and
$\sum \epsilon_{i} \mathcal{B}(\eta_c\to X_i)$. The systematic uncertainties from different sources
are listed in Table~\ref{sys_summary}. All sources are treated as uncorrelated, so the total
systematic uncertainty is obtained by summing them in quadrature. The following subsections describe
the procedures and assumptions that led to these estimates of the uncertainties.

\subsection{Luminosity}

The integrated luminosity is measured using Bhabha events, with an uncertainty of
$1.0\%$~\cite{Ablikim:2015nan}.

\subsection{Signal shape}

In the fit procedure, a discrepancy in the mass resolution between data and MC, as well as choices of
background shapes and fit range introduce uncertainties on the results. Since the statistical
fluctuation is large in the data sets, we cannot obtain a stable and reasonable estimation by simply
comparing two fits with different choices. To avoid the influence of statistical fluctuations, ensembles of simulated
data samples (toy MC samples) are generated according to an alternative fit model with the same statistics as data, then
fitted by the nominal model and the alternative model. These trials are performed 500 times, and the
deviation of mean values in the two trials is taken as the systematic uncertainty. The data samples taken at
$\sqrt{s}= 4.226$, 4.258, 4.358, and 4.416\,GeV are used to obtain an average uncertainty.

A discrepancy in mass resolution and mass scale between data and MC simulation affects the fit result. To estimate
this uncertainty, the signal shape is smeared and shifted by convolving it with a Gaussian function with a mean value of
$-1.2$\,MeV and standard deviation of $0.04$\,MeV, which are obtained from the study of a 
control sample of $e^+e^-\to\eta J/\psi$. Toy MC samples are generated according to the smeared MC
shape and fitted with a smeared and unsmeared signal shape.  The average deviation determined from
the four data samples is 7.5\% and is taken as systematic uncertainty.

\subsection{Background shape}

Similarly, to estimate the uncertainty due to the background shape, a sum of signal shape and a
second-order polynomial function with parameters determined from the fit on data is used to generate
toy MC, then the toy MC samples are fitted by models with a first-order and a second-order
polynomial background, respectively. The average deviation from the four data samples is found to be
6.3\% and is taken as systematic uncertainty.

\subsection{Fitting range}

The systematic uncertainty for the fit range is determined by varying the fit ranges randomly for 400
times. The standard deviation of the fit results is taken as systematic uncertainty, which is
determined to be 2.8\% from the four data samples.

\subsection{\boldmath $\mathcal{B}({h}_c\to\gamma \eta_c)\mathcal{B}(\eta\to\gamma \gamma)$ }

The branching fraction of $h_c\to\gamma\eta_c$ is taken from Ref.~\cite{Ablikim:2010rc}. The uncertainty
in this measurement is $15.7\%$ and the uncertainty of $\mathcal{B}(\eta\to\gamma \gamma)$ is $0.5\%$~\cite{Olive:2016xmw}. These uncertainties propagate to the cross section measurement.

\subsection{ISR correction}

To obtain the ISR correction factor, the energy dependent cross section is parameterized with the
sum of 3 coherent BW functions fitted to the cross sections measured in this analysis and the CLEO
value at 4.17\,GeV~\cite{CLEO:2011aa}.  The uncertainty of the input cross section is estimated by
two alternative models. First, the energy-dependent cross sections are fitted with a sum of BW and
a second order polynomial function. Second, the cross sections are fitted with a second order
polynomial function only. The maximum difference in ISR correction factor and detection efficiency
among these hypotheses is taken as systematic uncertainty due to the ISR correction.

\subsection{Vacuum polarization correction}

To investigate the uncertainty due to the vacuum polarization factor, we use two available VP
parameterisations~\cite{Eidelman:1995ny, Actis:2010gg}. The difference between them is 0.3\% and is
taken as the systematic uncertainty.

\subsection{\boldmath $\sum\limits_{i} \epsilon_{i} \mathcal{B}(\eta_c\to X_i)$ }

The branching ratios $\mathcal{B}(\eta_c\to X_i)$ are taken from BESIII
measurements~\cite{Ablikim:2012ur}, and the uncertainty of each channel is given in
Table~\ref{tab:sys_eff}. The systematic uncertainties associated with the efficiency include many items:
tracking, photon and PID efficiency, $K^0_{S}$, $\pi^0$, $\eta$ and $\eta_{c}$ reconstruction,
kinematic fit, cross feed and size of the MC sample. The procedure to estimate each item is
described below, and the results are also listed in Table~\ref{tab:sys_eff}.

\begin{itemize}

\item{Charged track, photon reconstruction and PID efficiencies }

  Both the tracking and PID efficiency uncertainties for charged tracks from the interaction point are determined to be 1\% per track, using the control samples of $J/\psi\to \pi^{+}\pi^{-}\pi^{0}$, $J/\psi\to p\bar{p}\pi^{+}\pi^{-}$
  and $J/\psi\to K_{S}^{0}K^{+}\pi^{-}+c.c.$~\cite{Ablikim:2011kv}.  The uncertainty due to the
  reconstruction of photons is 1\% per photon and it is determined
  from studies of $e^{+}e^{-}\to\gamma\mu^{+}\mu^{-}$ control samples~\cite{Prasad:2015bra}.

\item{$K_{S}^0$ efficiency}

  The uncertainty caused by $K_{S}^0$ reconstruction is studied with the processes
  $J/\psi\to K^{*\pm}K^{\mp}$ and $J/\psi\to \phi K^{0}_{S}K^{\pm}\pi^{\mp}$. 
  The discrepancy of $K_{S}^0$ reconstruction efficiency between data and MC simulation is found to be 1.2\%   and is taken as systematic uncertainty.

\item{$\eta$/$\pi^0$ efficiency}

  To estimate the uncertainty due to the resolution difference in $M(\gamma\gamma)$ between
  data and MC simulation in the $\eta$ and $\pi^0$ candidate selection, the MC shape of $\eta$
  ($\pi^0$) is smeared by convolving it with a Gaussian function that represents the discrepancy of
  resolution and is determined by the study of an $e^+e^-\to \eta J/\psi$ control sample.  The
  difference of reconstruction efficiencies with and without smearing is taken as systematic uncertainty.

\item{$\eta_c$ decay model}

  We use phase space to simulate $\eta_c$ decays in our analysis. To estimate the systematic
  uncertainty due to neglecting intermediate states in these decays, we study the intermediate
  states in $\eta_c$ decays from $\psi(3686)\to\gamma\eta_c,\eta_c \to X_i$ and generate MC samples
  accordingly. For channels with well-understood intermediate states, MC samples with these
  intermediate states are generated according to the relative branching ratios given by
  PDG~\cite{Olive:2016xmw}. The spreads of the efficiencies obtained from the phase-space and
  alternative MC samples are taken as the systematic uncertainties.

\item{$\eta_c$ line shape}

  The uncertainties of the $\eta_{c}$ line shape originate from the model of $\eta_c$ and the errors
  of its resonant parameters.  In the current MC generator, the $\eta_c$ line shape is described by a
  BW function. However, in $E$1 transitions $h_c\to\gamma\eta_c$ a cubic photon energy term with a
  damping term at higher energies is introduced to the signal shape because of the transition matrix
  element and phase space factor. To estimate this uncertainty, toy MC samples, generated according
  to the model that takes the $E$1 photon energy dependency into account, are analyzed to obtain the
  efficiency difference. The uncertainties due to the $\eta_c$ resonant parameters are considered by
  varying $m_{\eta_c}$ and $\Gamma_{\eta_c}$ in the MC simulation within their errors given by
  PDG~\cite{Olive:2016xmw}. The sum of these two items added in quadrature is taken as systematic
  uncertainty due to the $\eta_c$ line shape.

\item{Kinematic fit}

  For the signal MC samples, corrections to the track helix parameters and the corresponding
  covariance matrix for all charged tracks are made to obtain improved agreement between data and MC
  simulation~\cite{Ablikim:2012pg}. The difference between
  the obtained efficiencies with and without this correction is taken as the systematic uncertainty
  due to the kinematic fit.

\item{Cross feed}

  To check the contamination among the 16 decay modes of $\eta_c$, 40,000 MC events for each channel
  are used to test the event misjudgement.

\item{Size of the MC sample}

  The efficiency of each channel is obtained by MC simulation. The statistical uncertainty is
  calculated according to a binomial distribution.

\end{itemize}

In the fit procedure,
$\epsilon_{i} \mathcal{B}(\eta_c\to X_i) / \sum \epsilon_{i} \mathcal{B}(\eta_c\to X_i)$ is used to
constrain the strength among different $\eta_c$ decay modes, so the uncertainty from
$\epsilon_{i} \mathcal{B}(\eta_c\to X_i)$ will affect the fit results. In this case, we cannot
simply add the uncertainty from $\epsilon_{i} \mathcal{B}(\eta_c\to X_i)$ in quadrature with the other
uncertainties. To consider the uncertainties of $\epsilon_{i} \mathcal{B}(\eta_c\to X_i)$ and their
influence to the simultaneous fit, we change the $\epsilon_{i} \mathcal{B}(\eta_c\to X_i)$ within
their errors and refit the data sample. The change of the cross section with the new results is
taken as systematic uncertainty.

In this procedure, systematic uncertainties are divided into two categories: the correlated part,
which includes tracking, photon efficiency, PID efficiency, $\pi^{0}$/$\eta/K_{S}^{0}$ efficiency,
$\eta_{c}$ line shape and kinematic fit, and the uncorrelated part, which includes the $\eta_{c}$ decay
mode, cross feed, MC samples and $\mathcal{B}(\eta_{c}\to X)$. These uncertainties are assumed to be
distributed according to a Gaussian distribution. The uncertainties of the correlated part are changed dependently (increasing
or decreasing at the same time for all channels), while the uncertainties of the uncorrelated part
are changed independently.  We change the uncertainties (both correlated and uncorrelated parts) with
a Gaussian constraint and refit the data set 500 times. The cross sections calculated with these
trials are fitted with a Gaussian function, whose standard deviation is taken as systematic
uncertainty.  To obtain a conservartive estimation, the maximum deviation of $16.7\%$ from the data samples
at $\sqrt{s}= 4.226$, 4.258, 4.358 and 4.416\,GeV is adopted as systematic uncertainty from
$\sum_{i}\epsilon_{i} \times \mathcal{B}(\eta_c\to X_i)$ for all the data sets.

\begin{table}[htbp]
  \centering
   \caption{Summary of systematic uncertainties on $\sigma^{\rm{B}}(e^+e^-\to \eta h_c)$ (in \%) at $\sqrt{s}=4.226$\,GeV.}
  \begin{tabular}{lc}
  \hline
  \hline
  Sources    \hspace{5em}                                        & uncertainty in $\sigma^{\rm{B}}$  \\
  \hline
 A. Luminosity                                                          & $1.0$                         \\
 B. Signal shape                                                        & $7.5$                         \\
 C. Background shape                                                    & $6.3$                         \\
 D. Fitting range                                                       & $2.8$                         \\
 E. $\mathcal{B}({h}_c\to\gamma \eta_c)\mathcal{B}(\eta\to\gamma \gamma)$ & $15.7$                       \\
 F. ISR correction                                                      & $13.9$                         \\
 G. VP correction                                                       & $0.3$                         \\
 H. $\Sigma_{i}\epsilon_{i} \times \mathcal{B}(\eta_c\to X_i)$          & $16.7$                         \\
 \hline
 Total                                                               & $28.7$                          \\
  \hline
  \hline
    \end{tabular}
    \label{sys_summary}
\end{table}

\begin{table*}[tbhp]
  \caption{Systematic uncertainties (in \%) for $ \epsilon_{i} \mathcal{B}(\eta_c\to X_i)$ for each $\eta_c$ exclusive decay channel.}
  \begin{center}
    \vspace{0.5cm}
    \begin{tabular}{lcccccccc }
      \hline
      \hline
           Sources &  \large{$^{p \bar{p}}$} &  \large{$_{2(\pi^+ \pi^-)}$} &  \large{$^{2(K^+ K^-)}$} &  \large{$_{K^+ K^- \pi^+ \pi^-}$} &  \large{$^{p\bar{p}\pi^+ \pi^-}$} &  \large{$_{3(\pi^+ \pi^-)}$} &  \large{$^{K^+ K^- 2(\pi^+ \pi^-)}$} &  \large{$_{K^+ K^- \pi^0}$}   \\ \hline
           Tracking eff.                         & 2.0 & 4.0 & 4.0 & 4.0 & 4.0 & 6.0  & 6.0   & 2.0          \\
           Photon eff.                           & 3.0 & 3.0 & 3.0 & 3.0 & 3.0 & 3.0  & 3.0   & 5.0          \\
           PID                                   & 2.0 & 4.0 & 4.0 & 4.0 & 4.0 & 6.0  & 6.0   & 2.0          \\
           $K^0_S$ eff.                          &  -- &  -- &  -- &  -- &  -- &  --  &  --   &  --          \\
           $\pi^0$ eff.                          &  -- &  -- &  -- &  -- &  -- &  --  &  --   & 3.0          \\
           $\eta$ eff.                           & 1.7 & 2.0 & 2.6 & 1.7 & 1.7 & 1.7  & 2.0   & 1.3          \\
           $\eta_c$ decay model                  & 0.0 & 2.1 & 3.7 & 0.6 & 2.5 & 0.0  & 3.0   & 4.6          \\
           $\eta_c$ line shape                   & 6.2 & 6.2 & 6.2 & 6.2 & 6.2 & 6.2  & 5.0   & 5.0          \\
           Kinematic fit                         & 2.3 & 3.8 & 3.9 & 3.5 & 3.0 & 6.1  & 4.4   & 1.3          \\
           Cross feed                            & 0.0 & 0.7 & 0.0 & 2.0 & 0.0 & 0.0  & 0.0   & 0.0          \\
           MC sample                             & 0.9 & 1.2 & 1.7 & 1.4 & 1.4 & 1.6  & 2.1   & 1.4          \\
           $\mathcal{B}(\eta_c\to X_i)$          & 37.0& 22.0& 46.0& 26.0& 34.0& 28.0 & 54.0  & 23.0         \\
           \hline
           \hline
          \end{tabular} \label{tab:sys_eff}
  \end{center}
\end{table*}

\begin{table*}[tbhp]
  \begin{center}
    \vspace{0.5cm}
    \begin{tabular}{lcccccccc }
      \hline
      \hline
           Sources &  \large{$^{p \bar{p} \pi^0}$} &  \large{$_{K^0_{S} K^\pm \pi^\mp}$} &  \large{$^{K^0_{S} K^\pm \pi^\mp \pi^\pm \pi^\mp}$} &  \large{$_{\pi^+ \pi^- \eta}$} &  \large{$^{K^+ K^- \eta}$} &  \large{$ _{2(\pi^+ \pi^-) \eta}$} &  \large{$^{\pi^+ \pi^- \pi^0 \pi^0} $} &  \large{$_{2(\pi^+ \pi^- \pi^0)}$} \\ \hline
           Tracking eff.                         & 2.0 & 4.0 & 6.0 & 2.0 & 2.0 & 4.0  & 2.0   & 4.0      \\
           Photon eff.                           & 5.0 & 3.0 & 3.0 & 5.0 & 5.0 & 5.0  & 7.0   & 7.0      \\
           PID                                   & 2.0 & 4.0 & 6.0 & 2.0 & 2.0 & 4.0  & 2.0   & 4.0      \\
           $K^0_S$ eff.                          &  -- & 1.2 & 1.2 &  -- &  -- &  --  &  --   &  --      \\
           $\pi^0$ eff.                          & 2.3 &  -- &  -- &  -- &  -- &  --  & 3.1   & 1.5      \\
           $\eta$ eff.                           & 2.2 & 1.6 & 2.1 & 1.7 & 1.5 & 2.2  & 1.9   & 1.2      \\
           $\eta_c$ decay model                  & 5.8 & 2.5 & 5.2 & 5.5 & 8.1 & 0.0  & 0.1   & 0.5      \\
           $\eta_c$ line shape                   & 5.1 & 5.0 & 6.2 & 5.0 & 6.2 & 5.0  & 5.1   & 5.1      \\
           Kinematic fit                         & 2.1 & 2.4 & 1.7 & 1.1 & 2.5 & 1.8  & 0.5   & 2.9      \\
           Cross feed                            & 0.0 & 0.0 & 0.0 & 0.0 & 7.7 & 0.0  & 0.0   & 0.0      \\
           MC sample                             & 1.4 & 1.4 & 2.1 & 1.4 & 1.5 & 1.9  & 1.9   & 2.9      \\
           $\mathcal{B}(\eta_c\to X_i)$          & 38.0& 21.0& 28.0& 28.0& 54.0& 30.0 & 22.0  & 20.0     \\
           \hline
           \hline
          \end{tabular} \label{tab:sys2-total}
  \end{center}
\end{table*}

\section{Upper limit with systematic uncertainty}

For the data sets without significant $\eta h_c$ signals observed, an  upper limit at the 90\% C.L. on
the cross section is set using a Bayesian method, assuming a flat prior in $\sigma$. In this method, the probability density function of the measured
cross section $\sigma$, $P(\sigma)$, is determined using a maximum likelihood fit. The 90\%
confidence limit (L) is then calculated by solving the equation

\begin{equation}\label{eq3}
0.1=\int_{\rm{L}}^{\infty} P(\sigma) d\sigma.
\end{equation}

To include multiplicative systematics, $P(\sigma)$ is convolved with a probability distribution
function of sensitivity, which refers to the denominator of Eq.~(\ref{eq1}) and is assumed to be a
Gaussian with central value $\hat{S}$ and standard deviation $\sigma_{s}$~\cite{KStenson}:

\begin{equation}\label{eq4}
P^{\prime}(\sigma)=\int_0^{\infty} P\left(\frac{S}{\hat{S}}\sigma\right) \exp\left[\frac{-(S-\hat{S})^2}{2\sigma_{s}^2}\right]dS. 
\end{equation}

Here, $P(\sigma)$ is the likelihood distribution obtained from the fit and parameterized as double
Gaussian. By integrating $P^{\prime}(\sigma)$ we obtain the $90\%$ C.L.\ upper limit taking the systematic uncertainties
into account.

\section{\boldmath Results and discussion}

In this study, the Born cross section and its upper limits of $e^{+}e^{-}\to\eta h_{c}$ are measured
with statistical and systematical uncertainties at c.m.\ energies from 4.085 to 4.600\,GeV, and the results
are listed in Table~\ref{Finalresults}. Clear signals of $e^{+}e^{-}\to\eta h_{c}$ are observed at
$\sqrt{s}=4.226$\,GeV for the first time. The Born cross section is measured to be
$(9.5^{+2.2}_{-2.0} \pm 2.7)$\,pb. We also observe evidence for the signal process at $\sqrt{s}=4.358$\,GeV with a cross
section of $(10.0^{+3.1}_{-2.7} \pm 2.6)$\,pb. For the other c.m.\ energies considered, no significant signals are
found, and upper limits on the cross section at the 90\% C.L. are determined.
The cross sections measured in this analysis and CLEO~\cite{CLEO:2011aa} are modeled with a coherent sum of
three BW functions (as shown in Fig.~\ref{FitBESandBELLE}) to calculate the ISR correction
factors. 

Comparing with the process $e^+e^-\to \eta J/\psi$~\cite{Ablikim:2015xhk}, if we suppose both processes come from higher mass vector charmonia, the ratio $\Gamma(\psi \to \eta h_{c} )/\Gamma(\psi \to \eta J/\psi )$ is determined to be $0.20\pm0.07$ and $1.79\pm0.84$ at $\sqrt{s}=4.23$\,GeV and $4.36$\,GeV, respectively. These results are larger than theoretical expectation: $\Gamma(\psi(4160) \to \eta h_{c} )/\Gamma(\psi(4160) \to \eta J/\psi ) = 0.07887$ and $\Gamma(\psi(4415) \to \eta h_{c} )/\Gamma(\psi(4415) \to \eta J/\psi ) = 0.06736$~\cite{Muhammad}.

Comparing with the cross section of $e^+e^-\to \pi^+ \pi^- h_c$~\cite{BESIII:2016adj},
we find that the cross section of $e^+e^-\to \eta h_c$ is smaller. But due to the limited statistics we cannot determine the line shape of c.m.~energy dependent cross section precisely.

\acknowledgments

The BESIII collaboration thanks the staff of BEPCII and the IHEP computing center for their strong support. This work is supported in part by National Key Basic Research Program of China under Contract No. 2015CB856700; National Natural Science Foundation of China (NSFC) under Contracts Nos. 11235011, 11322544, 11335008, 11425524, 11635010; the Chinese Academy of Sciences (CAS) Large-Scale Scientific Facility Program; the CAS Center for Excellence in Particle Physics (CCEPP); the Collaborative Innovation Center for Particles and Interactions (CICPI); Joint Large-Scale Scientific Facility Funds of the NSFC and CAS under Contracts Nos. U1232201, U1332201, U1532257, U1532258; CAS under Contracts Nos. KJCX2-YW-N29, KJCX2-YW-N45; 100 Talents Program of CAS; National 1000 Talents Program of China; INPAC and Shanghai Key Laboratory for Particle Physics and Cosmology; German Research Foundation DFG under Contracts Nos. Collaborative Research Center CRC 1044, FOR 2359; Istituto Nazionale di Fisica Nucleare, Italy; Koninklijke Nederlandse Akademie van Wetenschappen (KNAW) under Contract No. 530-4CDP03; Ministry of Development of Turkey under Contract No. DPT2006K-120470; National Natural Science Foundation of China (NSFC) under Contract No. 11505010; The Swedish Resarch Council; U. S. Department of Energy under Contracts Nos. DE-FG02-05ER41374, DE-SC-0010118, DE-SC-0010504, DE-SC-0012069; U.S. National Science Foundation; University of Groningen (RuG) and the Helmholtzzentrum fuer Schwerionenforschung GmbH (GSI), Darmstadt; WCU Program of National Research Foundation of Korea under Contract No. R32-2008-000-10155-0.


\begin{thebibliography}{99}


\bibitem{Ablikim:2007gd} 
  M.~Ablikim {\it et al.} [BES Collaboration],
  eConf C {\bf 070805}, 02 (2007)
  [Phys.\ Lett.\ B {\bf 660}, 315 (2008)];\\
  T.~Barnes, S.~Godfrey and E.~S.~Swanson,
  Phys.\ Rev.\ D {\bf 72}, 054026 (2005).


\bibitem{Aubert:2005rm} 
  B.~Aubert {\it et al.} [BaBar Collaboration],
  Phys.\ Rev.\ Lett.\  {\bf 95}, 142001 (2005).

\bibitem{He:2006kg} 
  Q.~He {\it et al.} [CLEO Collaboration],
  Phys.\ Rev.\ D {\bf 74}, 091104 (2006).


\bibitem{Yuan:2007sj} 
  C.~Z.~Yuan {\it et al.} [Belle Collaboration],
  Phys.\ Rev.\ Lett.\  {\bf 99}, 182004 (2007).


\bibitem{Ablikim:2015tbp} 
  M.~Ablikim {\it et al.} [BESIII Collaboration],
  Phys.\ Rev.\ Lett.\  {\bf 115}, 112003 (2015).


\bibitem{Aubert:2007zz} 
  B.~Aubert {\it et al.} [BaBar Collaboration],
  Phys.\ Rev.\ Lett.\  {\bf 98}, 212001 (2007).


\bibitem{Wang:2007ea}
  X.~L.~Wang {\it et al.}  [Belle Collaboration],
  Phys.\ Rev.\ Lett.\  {\bf 99}, 142002 (2007).




\bibitem{Choi:2003ue} 
  S.~K.~Choi {\it et al.} [Belle Collaboration],
  Phys.\ Rev.\ Lett.\  {\bf 91}, 262001 (2003).

\bibitem{Ablikim:2013mio} 
  M.~Ablikim {\it et al.} [BESIII Collaboration],
  Phys.\ Rev.\ Lett.\  {\bf 110}, 252001 (2013).



\bibitem{Liu:2013dau} 
  Z.~Q.~Liu {\it et al.} [Belle Collaboration],
  Phys.\ Rev.\ Lett.\  {\bf 110}, 252002 (2013).




\bibitem{Ablikim:2013xfr} 
  M.~Ablikim {\it et al.} [BESIII Collaboration],
  Phys.\ Rev.\ Lett.\  {\bf 112}, 022001 (2014).




\bibitem{Ablikim:2015gda} 
  M.~Ablikim {\it et al.} [BESIII Collaboration],
  Phys.\ Rev.\ Lett.\  {\bf 115}, 222002 (2015).






\bibitem{Ablikim:2013wzq}
  M.~Ablikim {\it et al.} [BESIII Collaboration],
  Phys.\ Rev.\ Lett.\  {\bf 111}, 242001 (2013).


\bibitem{Ablikim:2013emm} 
  M.~Ablikim {\it et al.} [BESIII Collaboration],
  Phys.\ Rev.\ Lett.\  {\bf 112}, 132001 (2014).



\bibitem{Ablikim:2014dxl}
  M.~Ablikim {\it et al.} [BESIII Collaboration],
  Phys.\ Rev.\ Lett.\  {\bf 113}, 212002 (2014).



\bibitem{Ablikim:2015vvn} 
  M.~Ablikim {\it et al.} [BESIII Collaboration],
  Phys.\ Rev.\ Lett.\  {\bf 115}, 182002 (2015).


\bibitem{Close:2005iz}
  F.~E.~Close and P.~R.~Page,
  Phys.\ Lett.\ B {\bf 628}, 215 (2005).

\bibitem{Zhu:2005hp}
  S.~-L.~Zhu,
  Phys.\ Lett.\ B {\bf 625}, 212 (2005);\\
  E.~Kou and O.~Pene,
  Phys.\ Lett.\ B {\bf 631}, 164 (2005);\\
  X.~Q.~Luo and Y.~Liu,
  Phys.\ Rev.\ D {\bf 74}, 034502 (2006)
  [Phys.\ Rev.\ D {\bf 74}, 039902 (2006)].


\bibitem{Chen:2016ejo} 
  Y.~Chen, W.~F.~Chiu, M.~Gong, L.~C.~Gui and Z.~Liu,
  Chin.\ Phys.\ C {\bf 40}, 081002 (2016).



\bibitem{Ebert:2005nc}
  D.~Ebert, R.~N.~Faustov and V.~O.~Galkin,
  Phys.\ Lett.\ B {\bf 634}, 214 (2006).

\bibitem{Maiani:2005pe}
  L.~Maiani, V.~Riquer, F.~Piccinini and A.~D.~Polosa,
  Phys.\ Rev.\ D {\bf 72}, 031502 (2005).

\bibitem{TWChiu}
  T.~W.~Chiu {\it et al.} [TWQCD Collaboration],
  Phys.\ Rev.\ D {\bf 73}, 094510 (2006).

\bibitem{Liu:2005ay}
  X.~Liu, X.~-Q.~Zeng and X.~-Q.~Li,
  Phys.\ Rev.\ D {\bf 72}, 054023 (2005).

\bibitem{CFQiao}
  C.~F.~Qiao,
  Phys.\ Lett.\ B {\bf 639}, 263 (2006).

\bibitem{Yuan:2005dr}
  C.~Z.~Yuan, P.~Wang and X.~H.~Mo,
  Phys.\ Lett.\ B {\bf 634}, 399 (2006).



\bibitem{Chen:2015dig} 
  H.-X.~Chen, L.~Maiani, A.~D.~Polosa and V.~Riquer,
  Eur.\ Phys.\ J.\ C {\bf 75}, 550 (2015).


\bibitem{Padmanath:2015era} 
  M.~Padmanath, C.~B.~Lang and S.~Prelovsek,
  Phys.\ Rev.\ D {\bf 92}, 034501 (2015).



\bibitem{Bugg:2008wu} 
  D.~V.~Bugg,
  J.\ Phys.\ G {\bf 35}, 075005 (2008).



\bibitem{Chen:2011xk} 
  D.~Y.~Chen and X.~Liu,
  Phys.\ Rev.\ D {\bf 84}, 034032 (2011).



\bibitem{Wang:2013cya} 
  Q.~Wang, C.~Hanhart and Q.~Zhao,
  Phys.\ Rev.\ Lett.\  {\bf 111}, 132003 (2013).




\bibitem{Swanson:2014tra} 
  E.~S.~Swanson,
  Phys.\ Rev.\ D {\bf 91},  034009 (2015).





\bibitem{Wang:2013kra} 
  Q.~Wang, M.~Cleven, F.~K.~Guo, C.~Hanhart, U.~G.~Mei\ss{}ner, X.~G.~Wu and Q.~Zhao,
  Phys.\ Rev.\ D {\bf 89}, 034001 (2014).



\bibitem{Lebed:2016hpi} 
  R.~F.~Lebed, R.~E.~Mitchell and E.~S.~Swanson,
  Prog.\ Part.\ Nucl.\ Phys.\  {\bf 93}, 143 (2017).



\bibitem{CLEO:2011aa}
  T.~K.~Pedlar {\it et al.}  [CLEO Collaboration],
  Phys.\ Rev.\ Lett.\  {\bf 107}, 041803 (2011).


\bibitem{BESIII:2016adj} 
  M.~Ablikim {\it et al.} [BESIII Collaboration],
  Phys.\ Rev.\ Lett.\  {\bf 118}, 092002 (2017).

\bibitem{Voloshin:2004mh} 
  M.~B.~Voloshin,
  Phys.\ Lett.\ B {\bf 604}, 69 (2004).



\bibitem{Tamponi:2015xzb}
  U.~Tamponi {\it et al.} [Belle Collaboration],
  Phys.\ Rev.\ Lett.\  {\bf 115}, 142001 (2015).





\bibitem{Ablikim:2015nan}
  M.~Ablikim {\it et al.} [BESIII Collaboration],
  Chin.\ Phys.\ C {\bf 39}, 093001 (2015).


\bibitem{Ablikim:2015zaa} 
  M.~Ablikim {\it et al.} [BESIII Collaboration],
  Chin.\ Phys.\ C {\bf 40}, 063001 (2016).


\bibitem{ref:bes3}
  M.~Ablikim {\it et al.}  [BESIII Collaboration],
  Nucl.\ Instrum.\ Meth.\ A {\bf 614}, 345 (2010).



\bibitem{Agostinelli:2002hh}
  S.~Agostinelli {\it et al.} [GEANT4 Collaboration],
  Nucl.\ Instrum.\ Meth.\  A~{\bf 506}, 250 (2003); \\
  Geant4 version: v09-03p0; Physics List simulation engine: BERT; Physics List engine packaging library: PACK 5.5.

\bibitem{Allison:2006ve}
  J.~Allison {\it et al.},
  IEEE Trans.\ Nucl.\ Sci.\ {\bf 53}, 270 (2006).

\bibitem{ref:kkmc} S. Jadach , B. F. L. Ward and Z. Was, Comp. Phys. Commun. {\bf 130}, 260 (2000);
S. Jadach, B. F. L. Ward  and  Z. Was, Phys. Rev. D {\bf 63}, 113009~(2001).

\bibitem{ref:bes3gen} R. G. Ping, Chin.\ Phys.\ C {\bf 32}, 599~(2008);
  D.~J.~Lange, Nucl.\ Instrum.\ Meth.\ A {\bf 462}, 152 (2001).



\bibitem{Olive:2016xmw} 
  C.~Patrignani {\it et al.} [Particle Data Group],
  Chin.\ Phys.\ C {\bf 40}, 100001 (2016).


\bibitem{Chen:2000tv}
  J.~C.~Chen, G.~S.~Huang, X.~R.~Qi, D.~H.~Zhang and Y.~S.~Zhu,
  Phys.\ Rev.\ D {\bf 62}, 034003 (2000).




\bibitem{Sjostrand:2001yu}
 T.~Sj\"{o}strand {\it et al.},
  Comput.\ Phys.\ Commun.\  {\bf 191}, 159 (2015).


\bibitem{ref::ks0-reconstruction} M. Xu, {\it et al.}, Chin.\ Phys.\ C {\bf 33} (2009) 428.



\bibitem{Balossini:2006wc} 
  G.~Balossini, C.~M.~Carloni Calame, G.~Montagna, O.~Nicrosini and F.~Piccinini,
  Nucl.\ Phys.\ B {\bf 758}, 227 (2006).


\bibitem{Kuraev:1985hb}
  E.~A.~Kuraev and V.~S.~Fadin,
  Yad.\ Fiz.\  {\bf 41}, 733 (1985)
  [Sov.\ J.\ Nucl.\ Phys.\  {\bf 41}, 466 (1985)].



\bibitem{Eidelman:1995ny}
  S.~Eidelman and F.~Jegerlehner,
  Z.\ Phys.\ C {\bf 67}, 585 (1995).


\bibitem{Jegerlehner:2011mw} 
  F.~Jegerlehner,
  Nuovo Cim.\ C {\bf 034S1}, 31 (2011).



\bibitem{Ablikim:2010rc}
  M.~Ablikim {\it et al.}  [BESIII Collaboration],
  Phys.\ Rev.\ Lett.\  {\bf 104}, 132002 (2010).



\bibitem{Actis:2010gg}
  S.~Actis {\it et al.}  [Working Group on Radiative Corrections and Monte Carlo Generators for Low Energies Collaboration],
  Eur.\ Phys.\ J.\ C {\bf 66}, 585 (2010).


\bibitem{Ablikim:2012ur}
  M.~Ablikim {\it et al.}  [BESIII Collaboration],
  Phys.\ Rev.\ D {\bf 86}, 092009 (2012).



\bibitem{Ablikim:2011kv}
  M.~Ablikim {\it et al.} [BESIII Collaboration],
  Phys.\ Rev.\ D {\bf 83}, 112005 (2011).


\bibitem{Prasad:2015bra} 
  V.~Prasad, C.~Liu, X.~Ji, W.~Li, H.~Liu and X.~Lou,
  Physics {\bf 174}, 577 (2016).

\bibitem{Ablikim:2012pg}
  M.~Ablikim {\it et al.} [BESIII Collaboration],
  Phys.\ Rev.\ D {\bf 87}, 012002 (2013).

\bibitem{KStenson}K.~Stenson,
    arXiv:physics/0605236.


\bibitem{Ablikim:2015xhk} 
  M.~Ablikim {\it et al.} [BESIII Collaboration],
  Phys.\ Rev.\ D {\bf 91}, 112005 (2015).

\bibitem{Muhammad} 
   M.~N.~Anwar, Y.~Lu, B.~S.~Zou, arXiv:1612.05396.


\end{thebibliography}
\end{document}